\documentclass[prx,aps,twocolumn,superscriptaddress,balancelastpage]{revtex4-1}
\usepackage{latexsym,amssymb,bm,bbold,amsfonts,amstext,graphicx,bbm,bm,relsize,dsfont,times}
\usepackage[dvipsnames]{xcolor}
\usepackage{enumerate}  

\usepackage{amsmath} 
\usepackage{amsthm} 

\usepackage[unicode=true, breaklinks=false, pdfborder={0 0 1}, backref=false, colorlinks=true, linkcolor=blue, citecolor=blue]{hyperref}

\usepackage{subfiles}

\newcommand{\tr}{\mbox{tr}}

\def\bra#1{\langle{#1}|}
\def\ket#1{|{#1}\rangle}
\def\braket#1{\langle{#1}\rangle}

\def\BraVert{\egroup\,\mid\,\bgroup}

\newtheorem{definition}{Definition}

\newtheorem{theorem}[definition]{Theorem}

\newtheorem{remark}[definition]{Remark}

\usepackage{tikz}
\usepackage{xcolor}
\DeclareRobustCommand*\circled[1]{\tikz[baseline=(char.base),every node/.style={scale=0.35}]{
            \node[shape=circle,fill] (char) {{\fontfamily{phv}\selectfont{\Huge{{\textcolor{white}{#1}}}}}};}}

\begin{document}

\title{Tightening Quantum Speed Limits for Almost All States}

\author{Francesco Campaioli}
\email{francesco.campaioli@monash.edu}
\affiliation{School of Physics and Astronomy, Monash University, Victoria 3800, Australia}

\author{Felix A. Pollock}
\affiliation{School of Physics and Astronomy, Monash University, Victoria 3800, Australia}

\author{Felix C. Binder}
\affiliation{School of Physical \& Mathematical Sciences, Nanyang Technological University, 637371 Singapore, Singapore}

\author{Kavan Modi}
\affiliation{School of Physics and Astronomy, Monash University, Victoria 3800, Australia}

\date{\today}

\begin{abstract}
Conventional quantum speed limits perform poorly for mixed quantum states: They are generally not tight and often significantly underestimate the fastest possible evolution speed. To remedy this, for unitary driving, we derive two quantum speed limits that outperform the traditional bounds for almost all quantum states. Moreover, our bounds are significantly simpler to compute as well as experimentally more accessible. Our bounds have a clear geometric interpretation; they arise from the evaluation of the angle between generalized Bloch vectors.
\vspace{5pt}
\newline
\end{abstract} 
\maketitle

\makeatletter

{\bf Quantum speed limits (QSLs)} set fundamental bounds on the shortest time required to evolve between two quantum states~\cite{Anandan1990,Vaidman1992, Deffner2017}. The earliest derivation of minimal time of evolution was in 1945 by Mandelstam and Tamm~\cite{Mandelstam1945} with the aim of operationalising the famous (but oft misunderstood) \emph{time-energy uncertainty relations}~\cite{Aharonov1961, Eberly1973, Bauer1978, Uffink1985, Gislason1985} $\Delta t \geq \hbar / \Delta E$, relating the standard deviation of energy with the time it takes to go from one state to another. QSLs were originally derived for the unitary evolution of pure states~\cite{Fleming1973, Uffink1993, Margolus1998}; since then they have been generalized to the case of mixed states~\cite{Uhlmann1992b,Deffner2013,Zhang2014,Mondal2016}, non-unitary evolution~\cite{Deffner, DelCampo2013, Taddei2013}, and multi-partite systems~\cite{Giovannetti2004, Zander2007, Borras2008, Batle2005, Batle2006}. 

Extending their original scope, their significance has evolved from fundamental physics to practical relevance, defining the limits of the rate of information transfer~\cite{Bekenstein1981} and processing~\cite{Lloyd2000}, entropy production~\cite{Deffner2010}, precision in quantum metrology~\cite{Giovannetti2011} and time-scale of quantum optimal control~\cite{Caneva2009}. For example, in~\cite{Murphy2010}, the authors use QSLs to calculate the maximal rate of information transfer along a spin chain; similarly, Reich \textit{et al.} show that optimization algorithms and QSLs can be used together to achieve quantum control over a large class of physical systems~\cite{Reich2012}. In Refs.~\cite{Binder2015,Binder2016,Campaioli2017} QSLs are used to bound the charging power of non-degenerate multi-partite systems, which are treated as batteries. The latter results imply a significant speed advantage for entangling over local unitary driving of quantum systems, given the same external constraints.

Combining the Mandelstam-Tamm result with the results by Margolus and Levitin, along with elements of quantum state space geometry~\cite{Bengtsson2008}, leads to a \textit{unified QSL}~\cite{Levitin2009}. It bounds the shortest time required to evolve a (mixed) state $\rho$ to another state $\sigma$ by means of a unitary operator $U_t$ generated by some time-dependent Hamiltonian $H_t$
\begin{gather}
    \label{eq:QSL_mixed}
    T_{\mathcal{L}}(\rho,\sigma) = \hbar \frac{\mathcal{L}(\rho,\sigma)}{\min \{E,\Delta E\}},
\end{gather}
where $\mathcal{L} (\rho,\sigma) =\arccos(\mathcal{F}(\rho,\sigma))$ is the Bures angle, a measure of the distance between states $\rho$ and $\sigma$,  $\mathcal{F}(\rho,\sigma) =\tr{[\sqrt{\sqrt{\rho}\sigma\sqrt{\rho}}]}$ is the Uhlmann \emph{root} fidelity~\cite{Wootters1981, Uhlmann1992a}; $\rho_t = U_t \rho U_t^\dagger$, $E=(1/T)\int_0^T(\tr[\rho_t H_t]-h_t^{(0)}) \ dt$ is the average energy, with $h_t^{(0)}$ being the ground state energy of $H_t$; and $\Delta E = (1/T) \int_0^T \sqrt{\tr[\rho_t H_t^2]-\tr[\rho_t H_t]^2} \ dt$ the standard deviation~\cite{Deffner2013} ($\hbar=1$, here and in the following).

For pure states $\rho=\ket{\psi}\bra{\psi}$ and $\sigma=\ket{\phi}\bra{\phi}$, the Bures angle reduces to the Fubini-Study distance $d(\ket{\psi},\ket{\phi}) = \arccos{|\braket{\psi|\phi}|}$~\cite{Bengtsson2008,Fubini1904,Study1905}. Under this condition Eq.~\eqref{eq:QSL_mixed} is provably tight~\cite{Levitin2009}. An insightful geometric interpretation of QSLs for pure states (in a Hilbert space of any dimension) is that the geodesic connecting initial and final states lives on a \emph{complex projective line} $\mathbb{C}P^1$ (isomorphic to a 2-sphere $S^2$), defined by the linear combinations of $\ket{\psi}$ and $\ket{\phi}$~\cite{Bengtsson2008}. Any optimal Hamiltonian drives the initial state to the final one along an arc of a great circle on the sphere $S^2$ associated with the linear subspace of Hilbert space $\mathcal{H}$ generated by $\ket{\psi}$ and $\ket{\phi}$. In the case of mixed states, on the other hand, the speed limit induced by the Bures metric is in general not tight. 

In this Letter, we derive a tighter bound for the speed of unitary evolution. We propose the use of the angle between generalized Bloch vectors~\cite{Baird1963, Byrd2003} as a distance for those elements of state space that can be unitarily connected, and show that it induces an attainable bound for the unitary evolution of mixed qubits. However, as it turns out, this distance does not reduce to the Fubini-Study distance when pure states of dimension $N>2$ are considered. We thus introduce another distance that reduces to the Fubini-Study distance for pure states, and we derive a corresponding speed limit from it. Careful analysis of both newly introduced QSLs -- analytical for qubits and numerical for higher dimensions -- shows that they are tighter than the one derived from the Bures angle for the vast majority of mixed states. We conclude with a unified bound for the speed limit of unitary evolution.

\begin{figure}
\centering
\includegraphics[width=0.5\textwidth]{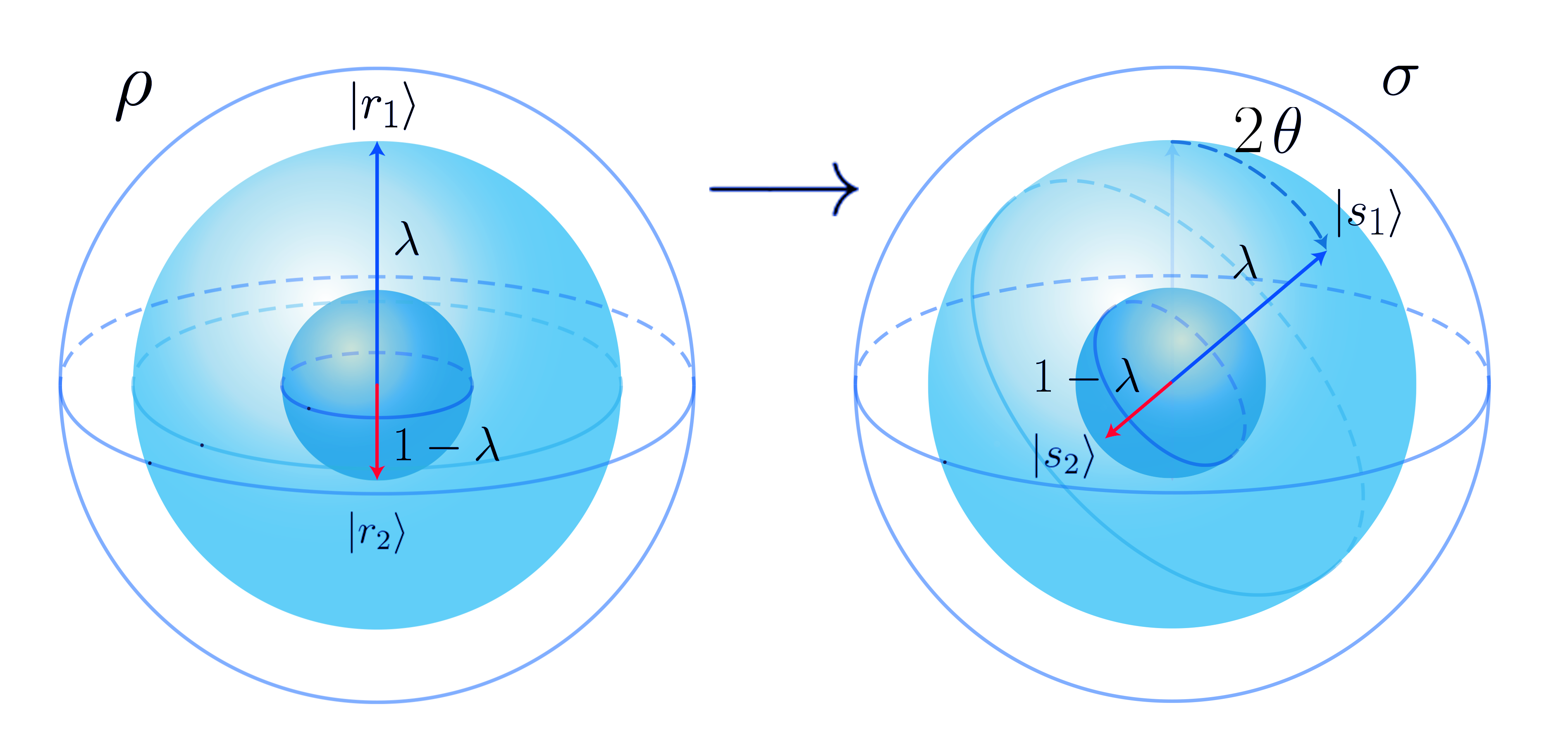}
\caption[width=0.5\textwidth]{(Color online.) Let $\rho$ and $\sigma$ be two mixed qubit states with the same spectrum, $\rho = \lambda \ket{r_1}\bra{r_1} + (1-\lambda) \ket{r_2}\bra{r_2}$, and $\sigma = \lambda \ket{s_1}\bra{s_1} + (1-\lambda) \ket{s_2}\bra{s_2}$, with $\lambda \in (0,1)$, excluding the maximally mixed state ($\lambda=1/2$), where $\{\ket{r_1},\ket{r_2}\}$ and $\{\ket{s_1},\ket{s_2}\}$ are two orthonormal bases. The problem of unitarily evolving $\rho$ to $\sigma$ can be mapped to evolving $\ket{r_1}$ to $\ket{s_1}$ (or, equivalently, $\ket{r_2}$ to $\ket{s_2}$). Eq.~\eqref{eq:QSL_mixed} is tight for pure states, thus any Hamiltonian that takes $\ket{r_1}$ to $\ket{s_1}$, will also take $\rho$ to $\sigma$ in the same time. For any Hamiltonian, with bounded standard deviation $\Delta E \le \mathcal{E}$, this time is bounded from below by $\theta/\mathcal{E}$, where $\theta = d(\ket{r_1},\ket{s_1})$ is the distance between $\ket{r_1}$ and $\ket{s_1}$, \emph{i.e} half of the angle between the vectors associated with $\ket{r_1}$ and $\ket{s_1}$. However, Eq.~\eqref{eq:QSL_mixed} for the same constraint on the Hamiltonian suggests that $T_{\mathcal{L}}=  \mathcal{L} (\rho,\sigma)/\mathcal{E}$, with $\mathcal{L}(\rho,\sigma)< \theta$ for every choice of $\lambda \neq 0,1$ [see Eq.~\eqref{eq:analytic_bures}], making the QSL unattainable for all mixed states.}
\label{fig:qubit_mixed}
\end{figure}

{\bf Attainability for mixed states --}
Let $\rho=\sum_i \lambda_i \ket{r_i}\bra{r_i}$ and $\sigma=\sum_i \lambda_i \ket{s_i}\bra{s_i}$ be two mixed states with the same spectrum. Let $\rho'=\sum_i \lambda'_i \ket{r_i}\bra{r_i}$ and $\sigma'=\sum_i \lambda'_i \ket{s_i}\bra{s_i}$ be another pair of mixed states with the same degeneracy structure as $\rho$ and $\sigma$, but different eigenvalues $\lambda'_i$. Any driving Hamiltonian that maps $\rho$ to $\sigma$ will map $\rho'$ to $\sigma'$ in the same amount of time, independent of their spectrum. On the other hand, the Bures angle is a continuous function of the spectrum of the mixed state, \textit{i.e.}, we could have $\mathcal{L} (\rho,\sigma) \approx 1$, while $\mathcal{L} (\rho',\sigma') \approx 0$. 
Even though the denominator of Eq.~\eqref{eq:QSL_mixed} may in principle also differ between these two scenarios due to its state-dependence~\footnote{Some authors have suggested quantifying the driving resource independently of the state, for instance in terms of norms of the driving Hamiltonian \cite{Uzdin2012,Binder2015, Binder2016, Campaioli2017, Deffner2017a, Russell2017}.}, that bound cannot be tight for the case of mixed states. This observation is particularly evident in the case of mixed qubits, as exemplified in Fig.~\ref{fig:qubit_mixed}.

The poor performance of the bound in Eq.~\eqref{eq:QSL_mixed} stems from the construction of the Bures distance for mixed states, which relies on purifying state $\rho_i$ to some $\ket{\psi_i}$ embedded in a larger Hilbert space  $\mathcal{H} \otimes \mathcal{H}_B$, such that $\mathrm{tr}_B [\ket{\psi_i} \bra{\psi_i}] = \rho_i$, where $\tr_B$ denotes the partial trace over $\mathcal{H}_B$. The distance $\mathcal{L}$ between two states $\rho_1$, $\rho_2 \in \mathcal{S}(\mathcal{H})$ is defined as the minimal Fubini-Study distance between the pure states $\ket{\psi_1}$, $\ket{\psi_2}$, where the \emph{minimum} is taken with respect to all possible unitary operations that act on the elements of $\mathcal{H}\otimes\mathcal{H_B}$. However, tracing over $\mathcal{H}_B$, in general, turns unitary dynamics between $\ket{\psi_1}$ and $\ket{\psi_2}$ into a non-unitary dynamics between $\rho_1$ and $\rho_2$~\cite{Breuer2002}. Consequently, the Bures metric does not necessarily select geodesics generated by unitary operations, even if $\rho_1$ and $\rho_2$ have the same spectrum.

The fact that Eq.~\eqref{eq:QSL_mixed} constitutes a loose bound for the speed of unitary evolution of mixed states is well known, and there are several proposals to tackle this problem \cite{Pires2016, Marvian2016, Mondal2016b}. In particular  Ref.~\cite{Pires2016} takes a geometric approach to obtain an infinite family of speed limits, while Ref.~\cite{Mondal2016b}  proposes a distance measure on the unitary orbit itself.

We now propose two distance measures for mixed states with the same fixed spectrum, that do not suffer from the problems outlined above. The corresponding QSLs outperform the bound in Eq.~\eqref{eq:QSL_mixed} and are much simpler to compute, as well as to experimentally measure. 

{\bf Generalized Bloch angle --}
Any mixed state $\rho \in \mathcal{S}(\mathcal{H})$ can be represented as
\begin{gather}
\label{eq:rho_bloch}
\rho=\frac{1}{N}\bigg(\mathbb{1}+\sqrt{\frac{N(N-1)}{2}}\; \bm{r}\cdot \bm{A} \bigg),
\end{gather}
where $N=\mathrm{dim}\mathcal{H}$, and $\bm{A} = (A_1,\dots, A_{N^2-1})$ is a set of operators that form a Lie algebra for SU($N$), such that $\tr[A_iA_j]=2\delta_{ij}$~\cite{Byrd2003}. The generalized Bloch vector $\bm{r}$ has to satisfy a set of relations in order to represent a state~\cite{Byrd2003}, such as $\bm{r}\cdot\bm{r}\leq 1$. 
We define the subset $\mathcal{S}_\Lambda(\mathcal{H}):=\{\rho\in\mathcal{S}(\mathcal{H}) : \mathrm{spec}(\rho)=\Lambda\}$ as the set of states with fixed spectrum $\Lambda$ that can be unitarily connected. 
The function
\begin{gather}
    \label{eq:gba}
    \Theta(\rho,\sigma)=  \arccos{\big(\hat{\bm{r}}\cdot\hat{\bm{s}}\big)},
\end{gather}
is a distance for the elements of $\mathcal{S}_\Lambda(\mathcal{H})$ for any fixed spectrum $\Lambda$, where $\hat{\bm{r}}$ and $\hat{\bm{s}}$ are the generalized Bloch vectors associated to states $\rho$ and $\sigma$, respectively, normalized for their length $\lVert \bm{r}\rVert_2 = \lVert \bm{s}\rVert_2$ (see proof of Theorem~\ref{th:qsl_gba}).
The angle $\Theta$ can be expressed as a function of $\rho$ and $\sigma$, independently from the chosen Lie algebra, $
    \Theta(\rho,\sigma)= \arccos{\big( (\tr[\rho\sigma] -1/N)/(\tr[\rho^2]-1/N) \big)}$,
using $\tr[\rho\sigma] = (1+(N-1)\bm{r}\cdot\bm{s})/N$. Note that the distance $\Theta(\rho,\sigma)$ does not depend on the basis chosen to represent the states, since the trace is basis-independent.

Our first result is a bound on the speed of unitary evolution for the elements of $\mathcal{S}_\Lambda(\mathcal{H})$ with fixed spectrum $\Lambda$, derived from the distance $\Theta$:
\begin{theorem}
\label{th:qsl_gba}
The minimal time required to evolve from state $\rho$ to state $\sigma$ by means of a unitary operation generated by the Hamiltonian $H_{t}$ is bounded from below by 
\begin{gather}
    \label{eq:qsl_gba}
    \begin{split}
    & T_\Theta (\rho,\sigma) =   \frac{\Theta(\rho,\sigma)}{ Q_\Theta},\;\textrm{where}\\
    &Q_\Theta = \frac{1}{T}\int_0^T dt \sqrt{\frac{2 \; \textrm{\emph{\tr}}[\rho_t^2 H_{t}^2-(\rho_t H_{t})^2]}{\textrm{\emph{\tr}}[\rho_t^2-\mathbb{1}/N^2]}}.
    \end{split}
\end{gather}
\end{theorem}
\noindent
\begin{proof} First, we prove that $\Theta$ is a distance. Let $\hat{\bm{r}}=\bm{r}/\lVert \bm{r} \rVert_2$, where $\lVert \bm{r} \rVert_2 = \sqrt{\bm{r} \cdot \bm{r}}$. Since $\hat{\bm{r}}\cdot\hat{\bm{s}} \in [-1,1] \Rightarrow \Theta(\rho,\sigma)\in[0,\pi]$, positivity holds. $\rho=\sigma \Rightarrow \hat{\bm{r}} = \hat{\bm{s}}$, thus $\Theta(\rho,\sigma) = 0$. $\Theta(\rho,\sigma)=0 \Rightarrow\bm{r}\cdot\bm{s}=\bm{r}\cdot\bm{r} =\bm{s}\cdot\bm{s}$, thus $\bm{r} =\bm{s}$ and so the identity of indiscernibility holds. Symmetry holds because $\hat{\bm{r}}\cdot\hat{\bm{s}} =\hat{\bm{s}}\cdot\hat{\bm{r}}$, thus $\Theta(\rho,\sigma) = \Theta(\sigma,\rho)$. Lastly, the triangle inequality holds, since $\hat{\bm{r}},\hat{\bm{s}}$ belong to the subset of $S^{N^2-1}_1$ ($N^2-1$ dimensional unit sphere) that satisfies the conditions for representing a state. It holds for all elements of $S^{N^2-1}_1$ that the angle between $\hat{\bm{r}}$ and $\hat{\bm{s}}$ is smaller than the sum of the angles between $\hat{\bm{r}}$, $\hat{\bm{q}}$ and $\hat{\bm{s}}$, $\hat{\bm{q}}$. 

To prove Eq.~\eqref{eq:qsl_gba},
we consider a state $\rho_{t+dt}=U_{t+dt:t} \ \rho_t \ U_{t+dt:t}^\dagger$ infinitesimally close to $\rho_t$, where $U_{t_2:t_1}$ is the unitary that maps $\rho_{t_1}$ to $\rho_{t_2}$.
We expand $U_{t+dt:t}$ for infinitesimal time $dt$ up to the second order 
to obtain $\rho_{t+dt} = \rho_t -i [H_t,\rho_t]dt -i [\partial H_t/\partial t,\rho_t]dt^2/2- \{H_t^2,\rho_t\}dt^2/2 +H_t\rho_t H_t dt^2 + \mathcal{O}(dt^3)$.
We then consider the inequality $\Theta(\rho,\sigma)\leq \int_0^T \Theta(\rho_t,\rho_{t+dt})$, that holds for any Hamiltonian $H_t$, where the equality may hold for optimal driving.
We calculate $\Theta(\rho_t,\rho_{t+dt}) = \arccos [1-dt^2(\tr[\rho_t^2 H_t^2]-\tr[(\rho_t H_t)^2])/(\tr[\rho_t^2]-1/N)]$, and expand $\arccos(1-c) = \sqrt{2}\sqrt{c}+\mathcal{O}(c)$ for small $c>0$, obtaining $\int_0^T\Theta(\rho_t,\rho_{t+dt}) = T Q_\Theta $, which leads to $\Theta(\rho,\sigma)\leq T Q_\Theta$, and thus to Eq.~\eqref{eq:qsl_gba}.
\end{proof}
\noindent
For pure states we would like Eq.~\eqref{eq:qsl_gba} to reduce to the unified bound~\eqref{eq:QSL_mixed}, obtained from the Fubini-Study metric~\cite{Fubini1904,Study1905}.
However, bound~\eqref{eq:qsl_gba} satisfies this requirement only for qubits:
\begin{remark}
\label{r:pure_limit}
Bound~\eqref{eq:qsl_gba} does not reduce to the QSL induced by the Fubini-Study metric for pure states, except for qubits  ($N=2$).
\end{remark}
\noindent
We give the proof in SM--\ref{s:pure_limit}. The reason why $\Theta$ does not conform with the Fubini-Study distance for pure states of arbitrary dimension  is that the group of rotations on the generalized Bloch vectors does not correspond to the group of unitary operators on states~\footnote{In the exceptional case of $N=2$, however, $\Theta$ does reduce to the Fubini-Study distance, since the set of all Bloch vectors forms a 2-sphere.}. When going from initial to final state, unitary evolution avoids the forbidden regions of the generalized Bloch sphere, whereas rotations would go straight through these regions, underestimating the distance between the considered states.

In order to derive a speed limit that conforms with the QSL for pure states regardless of the dimension of the system, we introduce the distance 
\begin{gather}
\label{eq:Theta_tilde}
    \Phi(\rho,\sigma)= \arccos \bigg(\sqrt{\frac{\mathrm{tr}[ \rho \sigma]}{\mathrm{tr} [\rho^2]}}\bigg),
\end{gather}
for the elements of $\mathcal{S}_\Lambda(\mathcal{H})$ for any fixed spectrum $\Lambda$, that reduces to the Fubini-Study distance for the case of pure states. If states of different purity were considered, neither $\Theta$ nor $\Phi$ would be distances, since the symmetry and triangle inequality properties would be lost.
As with $\Theta$, we derive a bound on the speed of unitary evolution from distance $\Phi$:
\begin{theorem}
\label{th:qsl_tilde}
The minimal time required to evolve from state $\rho$ to state $\sigma$ by means of a unitary operation generated by the Hamiltonian $H_t$ is bounded from below by 
\begin{gather}
    \label{eq:qsl_tilde}
    \begin{split}
    & T_{\Phi} (\rho,\sigma) =   \frac{\Phi(\rho,\sigma)}{ Q_\Phi},\; \textrm{where} \\
    & Q_\Phi = \frac{1}{T}\int_0^T dt \sqrt{\frac{\textrm{\emph{tr}}[\rho_t^2H_t^2-(\rho_t H_t)^2]}{\textrm{\emph{tr}}[\rho_t^2]}}.
    \end{split}
\end{gather}
\end{theorem}
\noindent 
The proof can be carried out using arguments similar to those for the proof of Theorem~\ref{th:qsl_gba}, see SM--\ref{s:proof_phi}. Remarkably, the bound expressed in Eq.~\eqref{eq:qsl_tilde} reduces to the Mandelstam-Tamm bound for pure states, since $\Phi$ reduces to the Fubini-Study distance and $Q_\Phi$ reduces to $\Delta E$~\footnote{To see this, note that the quantity {$\tr[(\rho_t H_t)^2]=\bra{a} H_t \ket{a}\bra{a} H_t\ket{a} = |\tr(\rho_t H_t)|^2$} for {$\rho=\ket{a}\bra{a}$}.}.

In contrast to bound~\eqref{eq:QSL_mixed}, the two QSLs derived here account for both the energetics of the dynamics and the purity of the driven state. The latter is accounted for by the denominators of $Q_\Theta$ and $Q_\Phi$, while the term in the numerators, $\sqrt{\tr[\rho_t^2 H_t^2-(\rho_t H_t)^2]}$, is a lower bound on the instantaneous standard deviation of the Hamiltonian $H_t$~\cite{Modi2016}.

It is worth highlighting that the bounds derived from $\Theta$ and $\Phi$ are significantly easier to compute than the one expressed in Eq.~\eqref{eq:QSL_mixed} for the case of mixed states, since no square root of density operators needs to be calculated, and thus no eigenvalue problem needs to be solved.
More specifically, in order to compute the Bures angle one needs to perform two matrix multiplications and two matrix square roots, whereas only two matrix multiplications are needed to compute $\Theta$ or $\Phi$ \footnote{The complexity of a matrix multiplication between two $N \times N$ matrices is equal to $O(N^{2.373})$, whereas the evaluation of the square root of such a matrix has complexity equal to $O(N^3)$ \cite{Frommer2010,Davie2013}}.
Accordingly, distances $\Theta$ and $\Phi$ can be experimentally estimated more efficiently than the Bures angle, which involves the evaluation of the root fidelity between the two considered states, and is harder to obtain than their overlap. The latter can be determined by means of a controlled-swap circuit~\cite{Ekert2002,Keyl2001}~\footnote{A similar experimental set up could be used to evaluate the bound derived by authors in Ref.~\cite{Mondal2016b}.}. Finally, not only are our bounds simpler to compute and measure, they also outperform  Eq.~\eqref{eq:QSL_mixed}, as we will show next.

\begin{figure}[h!]
    \centering
    \includegraphics[width=0.50\textwidth]{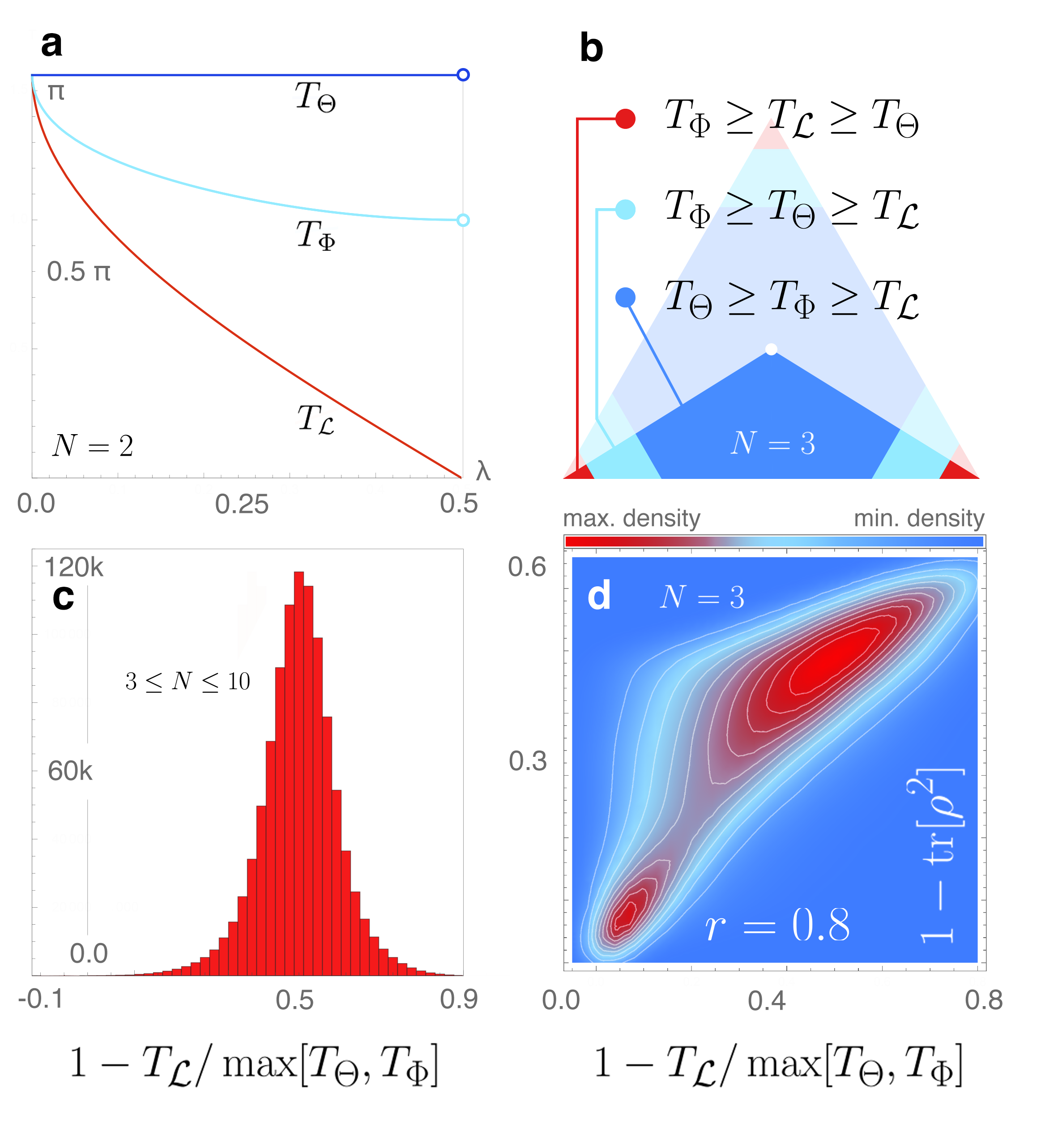}
    \caption[width=0.50\textwidth]{(Color online.) {\fontfamily{phv}\selectfont \textbf{a}} -- Bounds $T_{\mathcal{L}}$ (Eq.~\eqref{eq:QSL_mixed}), $T_\Phi$ (Eq.~\eqref{eq:qsl_tilde}), and $T_\Theta$ (Eq.~\eqref{eq:qsl_gba}), as a function of the eigenvalue $\lambda$, for two mixed and antipodal qubit states $\rho=\lambda \ket{r_1}\bra{r_1}+(1-\lambda)\ket{r_2}\bra{r_2}$ and $\sigma=\lambda \ket{r_2}\bra{r_2}+(1-\lambda)\ket{r_1}\bra{r_1}$. The unitary evolution is generated by the Hamiltonian $H=e^{i\varphi}\ket{r_1}\bra{r_2} + h.c.$. Bounds are symmetric with respect to $\lambda=1/2$. The same hierarchy holds for non-antipodal mixed qubit states. Bound $T_\Theta$ is always attainable.
    {\fontfamily{phv}\selectfont \textbf{b}}~--~For $N=3$ (qutrits), the hierarchy between the three bounds can be expressed with three regions of the polytope defined by the spectrum $\{\lambda_1, \lambda_2,\lambda_3\}$ of states $\rho$ and $\sigma$, as indicated in the legend. The corners of the triangle represent pure states, while its centre represent the maximally mixed state. The exact shape of the regions represented here reflects a specific choice of $H$, $\rho$ and $\sigma$, but similar features are common to those of any pair of states. For the case of qutrits, $T_{\mathcal{L}}$ is never larger than $\max[T_\Theta,T_\Phi]$ (see SM--\ref{s:qutrits} for more information).
    {\fontfamily{phv}\selectfont \textbf{c}}~--~Evaluation of $1-T_{\mathcal{L}}/\max[T_\Theta,T_\Phi]$ as a measure of the tightness of the new bounds, for $3\leq N\leq 10$, with a sample size of $10^6$ Haar random states and Hamiltonians. $T_{\mathcal{L}}$ can be larger than $\max[T_\Theta,T_\Phi]$, but only for 0.1 \% of the sampled states, and only with a difference of 1\% with respect to the largest of the new bounds.
    {\fontfamily{phv}\selectfont \textbf{d}}~--~Density plot of $10^5$ qutrit states, sampled approximately uniformly in terms of purity. The axes show numerical estimation of $1-T_{\mathcal{L}}/\max[T_\Theta,T_\Phi]$ (horizontal) and  $1-\tr[\rho^2]$ (vertical). We obtained the \emph{Pearson correlation coefficient} $r=0.8$. Bounds $T_\Theta$ and $T_{\mathcal{L}}$ coincide for pure states (bottom left), as shown analytically, and differ for increasingly mixed states (top right). This behaviour qualitatively extends to $N$-dimensional systems.}
    \label{fig:uniform}
\end{figure}

\vspace{5pt}
{\bf Attainability of new bounds --} 
We now study the bounds presented in Eqs.~\eqref{eq:qsl_gba} and~\eqref{eq:qsl_tilde}, and compare them to that in Eq.~\eqref{eq:QSL_mixed} for the same choice of initial state $\rho$ and Hamiltonian $H_t$. For the case of mixed qubits we calculate all three bounds analytically: Take $\rho=\lambda \ket{r_1} \bra{r_1} + (1-\lambda)\ket{r_2}\bra{r_2}$ as the initial state, and $H=e^{i\varphi} \ket{r_1} \bra{r_2} + h.c.$ as the Hamiltonian, where $\varphi\in[0,2\pi]$ is a phase. The chosen Hamiltonian generates the optimal unitary evolution for any choice of final state $\sigma=\lambda \ket{s_1} \bra{s_1} + (1-\lambda)\ket{s_2}\bra{s_2}$, for $\ket{s_1}=\cos\theta\ket{r_1} +e^{i\varphi}\sin\theta\ket{r_2}$.
The bounds read
\begin{align}
\label{eq:analytic_Theta}
&T_\Theta(\rho,\sigma)=\theta, \\
\label{eq:analytic_Phi}
&T_\Phi(\rho,\sigma)=\arccos\bigg({\sqrt{\frac{1+k^2\cos2\theta}{1-k^2}}}\;\bigg)\sqrt{\frac{1-k^2}{2 k^2}}, \\
\label{eq:analytic_bures}
&T_{\mathcal{L}}(\rho,\sigma) = \arccos \bigg( F_+(\theta,\lambda)+F_-(\theta,\lambda) \bigg),
\end{align}
where $F_\pm(\theta,\lambda)=\frac{1}{2}\sqrt{1+k^2c_{2\theta} \pm 2kc_{\theta}\sqrt{1-k^2s_\theta^2}}$, with $c_x = \cos x$, $s_x = \sin x$, and $k=1-2\lambda$.

Note that these bounds are independent of the relative phase $\varphi$, as we expect, and only depend on the distance $\theta=d(\ket{r_1}, \ket{s_1})$ between the basis elements, and on the value of $\lambda$. Bound $T_\Theta$ is tight and attainable and does not depend on the spectrum $\Lambda$. A simple plot of the bounds shows that $T_\Theta \geq T_\Phi \geq T_{\mathcal{L}}$ (see Fig.~\ref{fig:uniform} {\fontfamily{phv}\selectfont \textbf{a}}). The three bounds coincide for pure states $\lambda=0,1$ and for the trivial case of $\theta=0$.

In the general case of higher dimensions, we study the tightness of bounds $T_\Theta$ and $T_\Phi$ numerically (see Figs.~\ref{fig:uniform} {\fontfamily{phv}\selectfont \textbf{b}} and {\fontfamily{phv}\selectfont \textbf{c}}). Here, states are sampled so that their purity is approximately uniformly distributed between $1/N$ and $1$. The new bounds $T_\Theta$ and $T_{\mathcal{L}}$ coincide for pure states (as analytically shown above), and the difference between $\max[T_\Theta, T_\Phi]$ and $T_{\mathcal{L}}$ grows with decreasing purity (see Fig.~\ref{fig:uniform} {\fontfamily{phv}\selectfont \textbf{d}}).
Despite the fact that $T_\Theta$ and $T_\Phi$ are larger than $T_{\mathcal{L}}$ for the vast majority of cases, there are some exceptional regions where the latter can be larger than the new bounds, such as along some degenerate subspaces, which form a subset of measure zero of $\mathcal{S}_\Lambda(\mathcal{H})$. In the absence of a strict hierarchy between these bounds, we cast our main result in the form of a unified bound 
\begin{gather}
\label{eq:unified}
\begin{split}
    T_{\textrm{QSL}}(\rho,\sigma) = \max \big\{
    T_{\mathcal{L}},T_\Theta,T_\Phi
\big\},
\end{split}
\end{gather}
where $T_{\mathcal{L}}$, $T_\Theta$ and $T_\Phi$ are given by Eqs.~\eqref{eq:QSL_mixed},~\eqref{eq:qsl_gba} and~\eqref{eq:qsl_tilde}, respectively.

{\bf Conclusions --} 
In this Letter, we have addressed the problem of attainability of quantum speed limits for the unitary evolution of mixed states. We first showed that the conventional bound given in Eq.~\eqref{eq:QSL_mixed} is not generally tight for mixed states, because the Bures distance, defined as the minimal Fubini-Study distance on a dilated space, is not a suitable choice under the assumption of unitary evolution.

We have proposed two new distances between those elements of state space with the same spectrum, \emph{i.e.}, those that can be unitarily connected, and derived the corresponding QSLs. The first distance coincides with the angle between the generalized Bloch vectors and induces a tight and attainable speed limit for the case of mixed qubit states, but does not reduce to the unified bound in Eq.~\eqref{eq:QSL_mixed} for pure states of arbitrary dimension. The second distance is designed to conform for the case of pure states, while being as similar as possible to the generalized Bloch angle. These bounds arise from the properties of state space, when mixed states are represented as generalized Bloch vectors, providing thus a simple geometric interpretation. We have shown that the bounds obtained by these two distances are tighter than the conventional QSL given in Eq.~\eqref{eq:QSL_mixed} for the vast majority of states. Moreover, our new bounds are always easier to compute, as well as easier to measure experimentally. 

Beyond its fundamental relevance, our result provides a tighter, and hence more accurate bound on the rate of information transfer and processing in the presence of classical uncertainty. For instance, the computational speed of a quantum computer that works between mixed states would be bounded by Eq.~\eqref{eq:unified}, rather than Eq.~\eqref{eq:QSL_mixed}. The latter bound would  wrongly suggest that, in order to speed up computation, one could simply add noise, reducing the purity of the considered states, with the effect of reducing the time required to evolve between them. This paradoxical situation is now ruled out by our new bound, which demonstrates that in the proximity of maximally mixed states, the time required to perform any unitary evolution is finite and comparable to the time required to perform the evolution between pure states.

There is a natural trade off between the tightness of a QSL and its computational complexity. The ideas presented in this Letter open the door to finding a distance based on the explicit geometric structure of (mixed) state space. Such a distance would allow for the derivation of a QSL that is guaranteed to be tight, but at the same time easy to compute. It also remains open to apply the ideas developed here for the case of non-unitary dynamics. Such a generalisation would require modifying our proposed distances such that they accommodate changes in purity. Operationally meaningful QSLs for open dynamics would be of great practical importance to both theorists and experimentalists alike; however, developing them would require a careful analysis of the resource accounting implicit in the choice of different distances.

\vspace{5pt}
\begin{acknowledgments}
\noindent 
We kindly acknowledge A. K. Pati for historical details regarding the foundations of quantum speed limits and minimal evolution times, and G. Adesso, M. Bukov, S. Campbell, L. C. C\'{e}leri, B. Russell, and D. O. Soares-Pinto for the insightful comments to the first version of the letter. F. B. acknowledges support by the National Research Foundation of Singapore (Fellowship No. NRF-NRFF2016-02).
\end{acknowledgments}


\bibliography{speedlimitmixed.bbl}

\begin{thebibliography}{59}%
\makeatletter
\providecommand \@ifxundefined [1]{%
 \@ifx{#1\undefined}
}%
\providecommand \@ifnum [1]{%
 \ifnum #1\expandafter \@firstoftwo
 \else \expandafter \@secondoftwo
 \fi
}%
\providecommand \@ifx [1]{%
 \ifx #1\expandafter \@firstoftwo
 \else \expandafter \@secondoftwo
 \fi
}%
\providecommand \natexlab [1]{#1}%
\providecommand \enquote  [1]{``#1''}%
\providecommand \bibnamefont  [1]{#1}%
\providecommand \bibfnamefont [1]{#1}%
\providecommand \citenamefont [1]{#1}%
\providecommand \href@noop [0]{\@secondoftwo}%
\providecommand \href [0]{\begingroup \@sanitize@url \@href}%
\providecommand \@href[1]{\@@startlink{#1}\@@href}%
\providecommand \@@href[1]{\endgroup#1\@@endlink}%
\providecommand \@sanitize@url [0]{\catcode `\\12\catcode `\$12\catcode
  `\&12\catcode `\#12\catcode `\^12\catcode `\_12\catcode `\%12\relax}%
\providecommand \@@startlink[1]{}%
\providecommand \@@endlink[0]{}%
\providecommand \url  [0]{\begingroup\@sanitize@url \@url }%
\providecommand \@url [1]{\endgroup\@href {#1}{\urlprefix }}%
\providecommand \urlprefix  [0]{URL }%
\providecommand \Eprint [0]{\href }%
\providecommand \doibase [0]{http://dx.doi.org/}%
\providecommand \selectlanguage [0]{\@gobble}%
\providecommand \bibinfo  [0]{\@secondoftwo}%
\providecommand \bibfield  [0]{\@secondoftwo}%
\providecommand \translation [1]{[#1]}%
\providecommand \BibitemOpen [0]{}%
\providecommand \bibitemStop [0]{}%
\providecommand \bibitemNoStop [0]{.\EOS\space}%
\providecommand \EOS [0]{\spacefactor3000\relax}%
\providecommand \BibitemShut  [1]{\csname bibitem#1\endcsname}%
\let\auto@bib@innerbib\@empty
\bibitem [{\citenamefont {Anandan}\ and\ \citenamefont
  {Aharonov}(1990)}]{Anandan1990}%
  \BibitemOpen
  \bibfield  {author} {\bibinfo {author} {\bibfnamefont {J.}~\bibnamefont
  {Anandan}}\ and\ \bibinfo {author} {\bibfnamefont {Y.}~\bibnamefont
  {Aharonov}},\ }\href {\doibase 10.1103/PhysRevLett.65.1697} {\bibfield
  {journal} {\bibinfo  {journal} {Phys. Rev. Lett.}\ }\textbf {\bibinfo
  {volume} {65}},\ \bibinfo {pages} {1697} (\bibinfo {year}
  {1990})}\BibitemShut {NoStop}%
\bibitem [{\citenamefont {Vaidman}(1992)}]{Vaidman1992}%
  \BibitemOpen
  \bibfield  {author} {\bibinfo {author} {\bibfnamefont {L.}~\bibnamefont
  {Vaidman}},\ }\href {\doibase 10.1119/1.16940} {\bibfield  {journal}
  {\bibinfo  {journal} {Am. J. Phys.}\ }\textbf {\bibinfo {volume} {60}},\
  \bibinfo {pages} {182} (\bibinfo {year} {1992})}\BibitemShut {NoStop}%
\bibitem [{\citenamefont {Deffner}\ and\ \citenamefont
  {Campbell}(2017)}]{Deffner2017}%
  \BibitemOpen
  \bibfield  {author} {\bibinfo {author} {\bibfnamefont {S.}~\bibnamefont
  {Deffner}}\ and\ \bibinfo {author} {\bibfnamefont {S.}~\bibnamefont
  {Campbell}},\ }\href {https://arxiv.org/abs/1705.08023} {\  (\bibinfo {year}
  {2017})},\ \Eprint {http://arxiv.org/abs/1705.08023} {arXiv:1705.08023}
  \BibitemShut {NoStop}%
\bibitem [{\citenamefont {Mandelstam}\ and\ \citenamefont
  {Tamm}(1945)}]{Mandelstam1945}%
  \BibitemOpen
  \bibfield  {author} {\bibinfo {author} {\bibfnamefont {L.}~\bibnamefont
  {Mandelstam}}\ and\ \bibinfo {author} {\bibfnamefont {I.}~\bibnamefont
  {Tamm}},\ }in\ \href {\doibase 10.1007/978-3-642-74626-0_8} {\emph {\bibinfo
  {booktitle} {Sel. Pap.}}}\ (\bibinfo  {publisher} {Springer Berlin
  Heidelberg},\ \bibinfo {address} {Berlin, Heidelberg},\ \bibinfo {year}
  {1945})\ pp.\ \bibinfo {pages} {115--123}\BibitemShut {NoStop}%
\bibitem [{\citenamefont {Aharonov}\ and\ \citenamefont
  {Bohm}(1961)}]{Aharonov1961}%
  \BibitemOpen
  \bibfield  {author} {\bibinfo {author} {\bibfnamefont {Y.}~\bibnamefont
  {Aharonov}}\ and\ \bibinfo {author} {\bibfnamefont {D.}~\bibnamefont
  {Bohm}},\ }\href {\doibase 10.1103/PhysRev.122.1649} {\bibfield  {journal}
  {\bibinfo  {journal} {Phys. Rev.}\ }\textbf {\bibinfo {volume} {122}},\
  \bibinfo {pages} {1649} (\bibinfo {year} {1961})}\BibitemShut {NoStop}%
\bibitem [{\citenamefont {Eberly}\ and\ \citenamefont
  {Singh}(1973)}]{Eberly1973}%
  \BibitemOpen
  \bibfield  {author} {\bibinfo {author} {\bibfnamefont {J.~H.}\ \bibnamefont
  {Eberly}}\ and\ \bibinfo {author} {\bibfnamefont {L.~P.}\ \bibnamefont
  {Singh}},\ }\href {\doibase 10.1103/PhysRevD.7.359} {\bibfield  {journal}
  {\bibinfo  {journal} {Phys. Rev. D}\ }\textbf {\bibinfo {volume} {7}},\
  \bibinfo {pages} {359} (\bibinfo {year} {1973})}\BibitemShut {NoStop}%
\bibitem [{\citenamefont {Bauer}\ and\ \citenamefont
  {Mello}(1978)}]{Bauer1978}%
  \BibitemOpen
  \bibfield  {author} {\bibinfo {author} {\bibfnamefont {M.}~\bibnamefont
  {Bauer}}\ and\ \bibinfo {author} {\bibfnamefont {P.}~\bibnamefont {Mello}},\
  }\href {\doibase 10.1016/0003-4916(78)90223-3} {\bibfield  {journal}
  {\bibinfo  {journal} {Ann. Phys. (N. Y).}\ }\textbf {\bibinfo {volume}
  {111}},\ \bibinfo {pages} {38} (\bibinfo {year} {1978})}\BibitemShut
  {NoStop}%
\bibitem [{\citenamefont {Uffink}\ and\ \citenamefont
  {Hilgevoord}(1985)}]{Uffink1985}%
  \BibitemOpen
  \bibfield  {author} {\bibinfo {author} {\bibfnamefont {J.~B.~M.}\
  \bibnamefont {Uffink}}\ and\ \bibinfo {author} {\bibfnamefont
  {J.}~\bibnamefont {Hilgevoord}},\ }\href {\doibase 10.1007/BF00739034}
  {\bibfield  {journal} {\bibinfo  {journal} {Found. Phys.}\ }\textbf {\bibinfo
  {volume} {15}},\ \bibinfo {pages} {925} (\bibinfo {year} {1985})}\BibitemShut
  {NoStop}%
\bibitem [{\citenamefont {Gislason}\ \emph {et~al.}(1985)\citenamefont
  {Gislason}, \citenamefont {Sabelli},\ and\ \citenamefont
  {Wood}}]{Gislason1985}%
  \BibitemOpen
  \bibfield  {author} {\bibinfo {author} {\bibfnamefont {E.~A.}\ \bibnamefont
  {Gislason}}, \bibinfo {author} {\bibfnamefont {N.~H.}\ \bibnamefont
  {Sabelli}}, \ and\ \bibinfo {author} {\bibfnamefont {J.~W.}\ \bibnamefont
  {Wood}},\ }\href {\doibase 10.1103/PhysRevA.31.2078} {\bibfield  {journal}
  {\bibinfo  {journal} {Phys. Rev. A}\ }\textbf {\bibinfo {volume} {31}},\
  \bibinfo {pages} {2078} (\bibinfo {year} {1985})}\BibitemShut {NoStop}%
\bibitem [{\citenamefont {Fleming}(1973)}]{Fleming1973}%
  \BibitemOpen
  \bibfield  {author} {\bibinfo {author} {\bibfnamefont {G.~N.}\ \bibnamefont
  {Fleming}},\ }\href {\doibase 10.1007/BF02819419} {\bibfield  {journal}
  {\bibinfo  {journal} {Nuovo Cim. A}\ }\textbf {\bibinfo {volume} {16}},\
  \bibinfo {pages} {232} (\bibinfo {year} {1973})}\BibitemShut {NoStop}%
\bibitem [{\citenamefont {Uffink}(1993)}]{Uffink1993}%
  \BibitemOpen
  \bibfield  {author} {\bibinfo {author} {\bibfnamefont {J.}~\bibnamefont
  {Uffink}},\ }\href {\doibase 10.1119/1.17368} {\bibfield  {journal} {\bibinfo
   {journal} {Am. J. Phys.}\ }\textbf {\bibinfo {volume} {61}},\ \bibinfo
  {pages} {935} (\bibinfo {year} {1993})}\BibitemShut {NoStop}%
\bibitem [{\citenamefont {Margolus}\ and\ \citenamefont
  {Levitin}(1998)}]{Margolus1998}%
  \BibitemOpen
  \bibfield  {author} {\bibinfo {author} {\bibfnamefont {N.}~\bibnamefont
  {Margolus}}\ and\ \bibinfo {author} {\bibfnamefont {L.~B.}\ \bibnamefont
  {Levitin}},\ }\href {\doibase 10.1016/S0167-2789(98)00054-2} {\bibfield
  {journal} {\bibinfo  {journal} {Phys. D Nonlinear Phenom.}\ }\textbf
  {\bibinfo {volume} {120}},\ \bibinfo {pages} {188} (\bibinfo {year}
  {1998})}\BibitemShut {NoStop}%
\bibitem [{\citenamefont {Uhlmann}(1992{\natexlab{a}})}]{Uhlmann1992b}%
  \BibitemOpen
  \bibfield  {author} {\bibinfo {author} {\bibfnamefont {A.}~\bibnamefont
  {Uhlmann}},\ }\href {\doibase 10.1016/0375-9601(92)90555-Z} {\bibfield
  {journal} {\bibinfo  {journal} {Phys. Lett. A}\ }\textbf {\bibinfo {volume}
  {161}},\ \bibinfo {pages} {329} (\bibinfo {year}
  {1992}{\natexlab{a}})}\BibitemShut {NoStop}%
\bibitem [{\citenamefont {Deffner}\ and\ \citenamefont
  {Lutz}(2013{\natexlab{a}})}]{Deffner2013}%
  \BibitemOpen
  \bibfield  {author} {\bibinfo {author} {\bibfnamefont {S.}~\bibnamefont
  {Deffner}}\ and\ \bibinfo {author} {\bibfnamefont {E.}~\bibnamefont {Lutz}},\
  }\href {\doibase 10.1088/1751-8113/46/33/335302} {\bibfield  {journal}
  {\bibinfo  {journal} {J. Phys. A Math. Theor.}\ }\textbf {\bibinfo {volume}
  {46}},\ \bibinfo {pages} {335302} (\bibinfo {year}
  {2013}{\natexlab{a}})}\BibitemShut {NoStop}%
\bibitem [{\citenamefont {Zhang}\ \emph {et~al.}(2014)\citenamefont {Zhang},
  \citenamefont {Han}, \citenamefont {Xia}, \citenamefont {Cao},\ and\
  \citenamefont {Fan}}]{Zhang2014}%
  \BibitemOpen
  \bibfield  {author} {\bibinfo {author} {\bibfnamefont {Y.-J.}\ \bibnamefont
  {Zhang}}, \bibinfo {author} {\bibfnamefont {W.}~\bibnamefont {Han}}, \bibinfo
  {author} {\bibfnamefont {Y.-J.}\ \bibnamefont {Xia}}, \bibinfo {author}
  {\bibfnamefont {J.-P.}\ \bibnamefont {Cao}}, \ and\ \bibinfo {author}
  {\bibfnamefont {H.}~\bibnamefont {Fan}},\ }\href {\doibase 10.1038/srep04890}
  {\bibfield  {journal} {\bibinfo  {journal} {Sci. Rep.}\ }\textbf {\bibinfo
  {volume} {4}},\ \bibinfo {pages} {4890} (\bibinfo {year} {2014})}\BibitemShut
  {NoStop}%
\bibitem [{\citenamefont {Mondal}\ \emph {et~al.}(2015)\citenamefont {Mondal},
  \citenamefont {Datta},\ and\ \citenamefont {Sazim}}]{Mondal2016}%
  \BibitemOpen
  \bibfield  {author} {\bibinfo {author} {\bibfnamefont {D.}~\bibnamefont
  {Mondal}}, \bibinfo {author} {\bibfnamefont {C.}~\bibnamefont {Datta}}, \
  and\ \bibinfo {author} {\bibfnamefont {S.}~\bibnamefont {Sazim}},\ }\href
  {\doibase 10.1016/j.physleta.2015.12.015} {\bibfield  {journal} {\bibinfo
  {journal} {Phys. Lett. A}\ }\textbf {\bibinfo {volume} {380}},\ \bibinfo
  {pages} {689} (\bibinfo {year} {2015})}\BibitemShut {NoStop}%
\bibitem [{\citenamefont {Deffner}\ and\ \citenamefont
  {Lutz}(2013{\natexlab{b}})}]{Deffner}%
  \BibitemOpen
  \bibfield  {author} {\bibinfo {author} {\bibfnamefont {S.}~\bibnamefont
  {Deffner}}\ and\ \bibinfo {author} {\bibfnamefont {E.}~\bibnamefont {Lutz}},\
  }\href {\doibase 10.1103/PhysRevLett.111.010402} {\bibfield  {journal}
  {\bibinfo  {journal} {Phys. Rev. Lett}\ }\textbf {\bibinfo {volume} {111}},\
  \bibinfo {pages} {010402} (\bibinfo {year} {2013}{\natexlab{b}})}\BibitemShut
  {NoStop}%
\bibitem [{\citenamefont {del Campo}\ \emph {et~al.}(2013)\citenamefont {del
  Campo}, \citenamefont {Egusquiza}, \citenamefont {Plenio},\ and\
  \citenamefont {Huelga}}]{DelCampo2013}%
  \BibitemOpen
  \bibfield  {author} {\bibinfo {author} {\bibfnamefont {A.}~\bibnamefont {del
  Campo}}, \bibinfo {author} {\bibfnamefont {I.~L.}\ \bibnamefont {Egusquiza}},
  \bibinfo {author} {\bibfnamefont {M.~B.}\ \bibnamefont {Plenio}}, \ and\
  \bibinfo {author} {\bibfnamefont {S.~F.}\ \bibnamefont {Huelga}},\ }\href
  {\doibase 10.1103/PhysRevLett.110.050403} {\bibfield  {journal} {\bibinfo
  {journal} {Phys. Rev. Lett.}\ }\textbf {\bibinfo {volume} {110}},\ \bibinfo
  {pages} {050403} (\bibinfo {year} {2013})}\BibitemShut {NoStop}%
\bibitem [{\citenamefont {Taddei}\ \emph {et~al.}(2013)\citenamefont {Taddei},
  \citenamefont {Escher}, \citenamefont {Davidovich},\ and\ \citenamefont {{De
  Matos Filho}}}]{Taddei2013}%
  \BibitemOpen
  \bibfield  {author} {\bibinfo {author} {\bibfnamefont {M.~M.}\ \bibnamefont
  {Taddei}}, \bibinfo {author} {\bibfnamefont {B.~M.}\ \bibnamefont {Escher}},
  \bibinfo {author} {\bibfnamefont {L.}~\bibnamefont {Davidovich}}, \ and\
  \bibinfo {author} {\bibfnamefont {R.~L.}\ \bibnamefont {{De Matos Filho}}},\
  }\href {https://journals.aps.org/prl/abstract/10.1103/PhysRevLett.110.050402}
  {\bibfield  {journal} {\bibinfo  {journal} {Phys. Rev. Lett.}\ }\textbf
  {\bibinfo {volume} {110}},\ \bibinfo {pages} {050402} (\bibinfo {year}
  {2013})}\BibitemShut {NoStop}%
\bibitem [{\citenamefont {Giovannetti}\ \emph {et~al.}(2004)\citenamefont
  {Giovannetti}, \citenamefont {Lloyd},\ and\ \citenamefont
  {Maccone}}]{Giovannetti2004}%
  \BibitemOpen
  \bibfield  {author} {\bibinfo {author} {\bibfnamefont {V.}~\bibnamefont
  {Giovannetti}}, \bibinfo {author} {\bibfnamefont {S.}~\bibnamefont {Lloyd}},
  \ and\ \bibinfo {author} {\bibfnamefont {L.}~\bibnamefont {Maccone}},\ }\href
  {\doibase 10.1088/1464-4266/6/8/028} {\bibfield  {journal} {\bibinfo
  {journal} {J. Opt. B Quantum Semiclassical Opt.}\ }\textbf {\bibinfo {volume}
  {6}},\ \bibinfo {pages} {S807} (\bibinfo {year} {2004})}\BibitemShut
  {NoStop}%
\bibitem [{\citenamefont {Zander}\ \emph {et~al.}(2007)\citenamefont {Zander},
  \citenamefont {Plastino}, \citenamefont {Plastino},\ and\ \citenamefont
  {Casas}}]{Zander2007}%
  \BibitemOpen
  \bibfield  {author} {\bibinfo {author} {\bibfnamefont {C.}~\bibnamefont
  {Zander}}, \bibinfo {author} {\bibfnamefont {a.~R.}\ \bibnamefont
  {Plastino}}, \bibinfo {author} {\bibfnamefont {A.}~\bibnamefont {Plastino}},
  \ and\ \bibinfo {author} {\bibfnamefont {M.}~\bibnamefont {Casas}},\ }\href
  {\doibase 10.1088/1751-8113/40/11/020} {\bibfield  {journal} {\bibinfo
  {journal} {J. Phys. A Math. Theor.}\ }\textbf {\bibinfo {volume} {40}},\
  \bibinfo {pages} {2861} (\bibinfo {year} {2007})}\BibitemShut {NoStop}%
\bibitem [{\citenamefont {Borras}\ \emph {et~al.}(2008)\citenamefont {Borras},
  \citenamefont {Zander}, \citenamefont {Plastino}, \citenamefont {Casas},\
  and\ \citenamefont {Plastino}}]{Borras2008}%
  \BibitemOpen
  \bibfield  {author} {\bibinfo {author} {\bibfnamefont {A.}~\bibnamefont
  {Borras}}, \bibinfo {author} {\bibfnamefont {C.}~\bibnamefont {Zander}},
  \bibinfo {author} {\bibfnamefont {A.~R.}\ \bibnamefont {Plastino}}, \bibinfo
  {author} {\bibfnamefont {M.}~\bibnamefont {Casas}}, \ and\ \bibinfo {author}
  {\bibfnamefont {A.}~\bibnamefont {Plastino}},\ }\href {\doibase
  10.1209/0295-5075/81/30007} {\bibfield  {journal} {\bibinfo  {journal} {EPL}\
  }\textbf {\bibinfo {volume} {81}},\ \bibinfo {pages} {30007} (\bibinfo {year}
  {2008})}\BibitemShut {NoStop}%
\bibitem [{\citenamefont {Batle}\ \emph {et~al.}(2005)\citenamefont {Batle},
  \citenamefont {Casas}, \citenamefont {Plastino},\ and\ \citenamefont
  {Plastino}}]{Batle2005}%
  \BibitemOpen
  \bibfield  {author} {\bibinfo {author} {\bibfnamefont {J.}~\bibnamefont
  {Batle}}, \bibinfo {author} {\bibfnamefont {M.}~\bibnamefont {Casas}},
  \bibinfo {author} {\bibfnamefont {A.}~\bibnamefont {Plastino}}, \ and\
  \bibinfo {author} {\bibfnamefont {A.~R.}\ \bibnamefont {Plastino}},\ }\href
  {\doibase 10.1103/PhysRevA.72.032337} {\bibfield  {journal} {\bibinfo
  {journal} {Phys. Rev. A}\ }\textbf {\bibinfo {volume} {72}},\ \bibinfo
  {pages} {032337} (\bibinfo {year} {2005})}\BibitemShut {NoStop}%
\bibitem [{\citenamefont {Batle}\ \emph {et~al.}(2006)\citenamefont {Batle},
  \citenamefont {Casas}, \citenamefont {Plastino},\ and\ \citenamefont
  {Plastino}}]{Batle2006}%
  \BibitemOpen
  \bibfield  {author} {\bibinfo {author} {\bibfnamefont {J.}~\bibnamefont
  {Batle}}, \bibinfo {author} {\bibfnamefont {M.}~\bibnamefont {Casas}},
  \bibinfo {author} {\bibfnamefont {A.}~\bibnamefont {Plastino}}, \ and\
  \bibinfo {author} {\bibfnamefont {A.~R.}\ \bibnamefont {Plastino}},\ }\href
  {\doibase 10.1103/PhysRevA.73.049904} {\bibfield  {journal} {\bibinfo
  {journal} {Phys. Rev. A}\ }\textbf {\bibinfo {volume} {73}},\ \bibinfo
  {pages} {049904} (\bibinfo {year} {2006})}\BibitemShut {NoStop}%
\bibitem [{\citenamefont {Bekenstein}(1981)}]{Bekenstein1981}%
  \BibitemOpen
  \bibfield  {author} {\bibinfo {author} {\bibfnamefont {J.~D.}\ \bibnamefont
  {Bekenstein}},\ }\href {\doibase 10.1103/PhysRevLett.46.623} {\bibfield
  {journal} {\bibinfo  {journal} {Phys. Rev. Lett.}\ }\textbf {\bibinfo
  {volume} {46}},\ \bibinfo {pages} {623} (\bibinfo {year} {1981})}\BibitemShut
  {NoStop}%
\bibitem [{\citenamefont {Lloyd}(2000)}]{Lloyd2000}%
  \BibitemOpen
  \bibfield  {author} {\bibinfo {author} {\bibfnamefont {S.}~\bibnamefont
  {Lloyd}},\ }\href {\doibase 10.1038/35023282} {\bibfield  {journal} {\bibinfo
   {journal} {Nature}\ }\textbf {\bibinfo {volume} {406}},\ \bibinfo {pages}
  {1047} (\bibinfo {year} {2000})}\BibitemShut {NoStop}%
\bibitem [{\citenamefont {Deffner}\ and\ \citenamefont
  {Lutz}(2010)}]{Deffner2010}%
  \BibitemOpen
  \bibfield  {author} {\bibinfo {author} {\bibfnamefont {S.}~\bibnamefont
  {Deffner}}\ and\ \bibinfo {author} {\bibfnamefont {E.}~\bibnamefont {Lutz}},\
  }\href {\doibase 10.1103/PhysRevLett.105.170402} {\bibfield  {journal}
  {\bibinfo  {journal} {Phys. Rev. Lett.}\ }\textbf {\bibinfo {volume} {105}},\
  \bibinfo {pages} {170402} (\bibinfo {year} {2010})}\BibitemShut {NoStop}%
\bibitem [{\citenamefont {Giovannetti}\ \emph {et~al.}(2011)\citenamefont
  {Giovannetti}, \citenamefont {Lloyd},\ and\ \citenamefont
  {Maccone}}]{Giovannetti2011}%
  \BibitemOpen
  \bibfield  {author} {\bibinfo {author} {\bibfnamefont {V.}~\bibnamefont
  {Giovannetti}}, \bibinfo {author} {\bibfnamefont {S.}~\bibnamefont {Lloyd}},
  \ and\ \bibinfo {author} {\bibfnamefont {L.}~\bibnamefont {Maccone}},\ }\href
  {\doibase 10.1038/nphoton.2011.35} {\bibfield  {journal} {\bibinfo  {journal}
  {Nat. Photonics}\ }\textbf {\bibinfo {volume} {5}},\ \bibinfo {pages} {222}
  (\bibinfo {year} {2011})}\BibitemShut {NoStop}%
\bibitem [{\citenamefont {Caneva}\ \emph {et~al.}(2009)\citenamefont {Caneva},
  \citenamefont {Murphy}, \citenamefont {Calarco}, \citenamefont {Fazio},
  \citenamefont {Montangero}, \citenamefont {Giovannetti},\ and\ \citenamefont
  {Santoro}}]{Caneva2009}%
  \BibitemOpen
  \bibfield  {author} {\bibinfo {author} {\bibfnamefont {T.}~\bibnamefont
  {Caneva}}, \bibinfo {author} {\bibfnamefont {M.}~\bibnamefont {Murphy}},
  \bibinfo {author} {\bibfnamefont {T.}~\bibnamefont {Calarco}}, \bibinfo
  {author} {\bibfnamefont {R.}~\bibnamefont {Fazio}}, \bibinfo {author}
  {\bibfnamefont {S.}~\bibnamefont {Montangero}}, \bibinfo {author}
  {\bibfnamefont {V.}~\bibnamefont {Giovannetti}}, \ and\ \bibinfo {author}
  {\bibfnamefont {G.~E.}\ \bibnamefont {Santoro}},\ }\href
  {https://journals.aps.org/prl/abstract/10.1103/PhysRevLett.103.240501}
  {\bibfield  {journal} {\bibinfo  {journal} {Phys. Rev. Lett.}\ }\textbf
  {\bibinfo {volume} {103}},\ \bibinfo {pages} {240501} (\bibinfo {year}
  {2009})}\BibitemShut {NoStop}%
\bibitem [{\citenamefont {Murphy}\ \emph {et~al.}(2010)\citenamefont {Murphy},
  \citenamefont {Montangero}, \citenamefont {Giovannetti},\ and\ \citenamefont
  {Calarco}}]{Murphy2010}%
  \BibitemOpen
  \bibfield  {author} {\bibinfo {author} {\bibfnamefont {M.}~\bibnamefont
  {Murphy}}, \bibinfo {author} {\bibfnamefont {S.}~\bibnamefont {Montangero}},
  \bibinfo {author} {\bibfnamefont {V.}~\bibnamefont {Giovannetti}}, \ and\
  \bibinfo {author} {\bibfnamefont {T.}~\bibnamefont {Calarco}},\ }\href
  {https://doi.org/10.1103/PhysRevA.82.022318} {\bibfield  {journal} {\bibinfo
  {journal} {Phys. Rev. A}\ }\textbf {\bibinfo {volume} {82}},\ \bibinfo
  {pages} {022318} (\bibinfo {year} {2010})}\BibitemShut {NoStop}%
\bibitem [{\citenamefont {Reich}\ \emph {et~al.}(2012)\citenamefont {Reich},
  \citenamefont {Ndong},\ and\ \citenamefont {Koch}}]{Reich2012}%
  \BibitemOpen
  \bibfield  {author} {\bibinfo {author} {\bibfnamefont {D.~M.}\ \bibnamefont
  {Reich}}, \bibinfo {author} {\bibfnamefont {M.}~\bibnamefont {Ndong}}, \ and\
  \bibinfo {author} {\bibfnamefont {C.~P.}\ \bibnamefont {Koch}},\ }\href
  {\doibase 10.1063/1.3691827} {\bibfield  {journal} {\bibinfo  {journal} {J.
  Chem. Phys.}\ }\textbf {\bibinfo {volume} {136}},\ \bibinfo {pages} {104103}
  (\bibinfo {year} {2012})}\BibitemShut {NoStop}%
\bibitem [{\citenamefont {Binder}\ \emph {et~al.}(2015)\citenamefont {Binder},
  \citenamefont {Vinjanampathy}, \citenamefont {Modi},\ and\ \citenamefont
  {Goold}}]{Binder2015}%
  \BibitemOpen
  \bibfield  {author} {\bibinfo {author} {\bibfnamefont {F.~C.}\ \bibnamefont
  {Binder}}, \bibinfo {author} {\bibfnamefont {S.}~\bibnamefont
  {Vinjanampathy}}, \bibinfo {author} {\bibfnamefont {K.}~\bibnamefont {Modi}},
  \ and\ \bibinfo {author} {\bibfnamefont {J.}~\bibnamefont {Goold}},\ }\href
  {\doibase 10.1088/1367-2630/17/7/075015} {\bibfield  {journal} {\bibinfo
  {journal} {New J. Phys.}\ }\textbf {\bibinfo {volume} {17}},\ \bibinfo
  {pages} {075015} (\bibinfo {year} {2015})}\BibitemShut {NoStop}%
\bibitem [{\citenamefont {Binder}(2016)}]{Binder2016}%
  \BibitemOpen
  \bibfield  {author} {\bibinfo {author} {\bibfnamefont {F.~C.}\ \bibnamefont
  {Binder}},\ }\emph {\bibinfo {title} {{Work, Heat, and Power of Quantum
  Processes}}},\ \href
  {https://ora.ox.ac.uk/objects/uuid:279871ea-3b2e-4baf-975c-1bd42b4961c3}
  {Ph.D. thesis},\ \bibinfo  {school} {University of Oxford} (\bibinfo {year}
  {2016})\BibitemShut {NoStop}%
\bibitem [{\citenamefont {Campaioli}\ \emph {et~al.}(2017)\citenamefont
  {Campaioli}, \citenamefont {Pollock}, \citenamefont {Binder}, \citenamefont
  {C{\'{e}}leri}, \citenamefont {Goold}, \citenamefont {Vinjanampathy},\ and\
  \citenamefont {Modi}}]{Campaioli2017}%
  \BibitemOpen
  \bibfield  {author} {\bibinfo {author} {\bibfnamefont {F.}~\bibnamefont
  {Campaioli}}, \bibinfo {author} {\bibfnamefont {F.~A.}\ \bibnamefont
  {Pollock}}, \bibinfo {author} {\bibfnamefont {F.~C.}\ \bibnamefont {Binder}},
  \bibinfo {author} {\bibfnamefont {L.}~\bibnamefont {C{\'{e}}leri}}, \bibinfo
  {author} {\bibfnamefont {J.}~\bibnamefont {Goold}}, \bibinfo {author}
  {\bibfnamefont {S.}~\bibnamefont {Vinjanampathy}}, \ and\ \bibinfo {author}
  {\bibfnamefont {K.}~\bibnamefont {Modi}},\ }\href {\doibase
  10.1103/PhysRevLett.118.150601} {\bibfield  {journal} {\bibinfo  {journal}
  {Phys. Rev. Lett.}\ }\textbf {\bibinfo {volume} {118}},\ \bibinfo {pages}
  {150601} (\bibinfo {year} {2017})}\BibitemShut {NoStop}%
\bibitem [{\citenamefont {Bengtsson}\ and\ \citenamefont
  {Życzkowski}(2008)}]{Bengtsson2008}%
  \BibitemOpen
  \bibfield  {author} {\bibinfo {author} {\bibfnamefont {I.}~\bibnamefont
  {Bengtsson}}\ and\ \bibinfo {author} {\bibfnamefont {K.}~\bibnamefont
  {Życzkowski}},\ }\href@noop {} {\emph {\bibinfo {title} {{Geometry of
  quantum states : an introduction to quantum entanglement}}}}\ (\bibinfo
  {publisher} {Cambridge University Press},\ \bibinfo {year} {2008})\ p.\
  \bibinfo {pages} {419}\BibitemShut {NoStop}%
\bibitem [{\citenamefont {Levitin}\ and\ \citenamefont
  {Toffoli}(2009)}]{Levitin2009}%
  \BibitemOpen
  \bibfield  {author} {\bibinfo {author} {\bibfnamefont {L.~B.}\ \bibnamefont
  {Levitin}}\ and\ \bibinfo {author} {\bibfnamefont {T.}~\bibnamefont
  {Toffoli}},\ }\href
  {https://journals.aps.org/prl/abstract/10.1103/PhysRevLett.103.160502}
  {\bibfield  {journal} {\bibinfo  {journal} {Phys. Rev. Lett.}\ }\textbf
  {\bibinfo {volume} {103}},\ \bibinfo {pages} {160502} (\bibinfo {year}
  {2009})}\BibitemShut {NoStop}%
\bibitem [{\citenamefont {Wootters}(1981)}]{Wootters1981}%
  \BibitemOpen
  \bibfield  {author} {\bibinfo {author} {\bibfnamefont {W.~K.}\ \bibnamefont
  {Wootters}},\ }\href {\doibase 10.1103/PhysRevD.23.357} {\bibfield  {journal}
  {\bibinfo  {journal} {Phys. Rev. D}\ }\textbf {\bibinfo {volume} {23}},\
  \bibinfo {pages} {357} (\bibinfo {year} {1981})}\BibitemShut {NoStop}%
\bibitem [{\citenamefont {Uhlmann}(1992{\natexlab{b}})}]{Uhlmann1992a}%
  \BibitemOpen
  \bibfield  {author} {\bibinfo {author} {\bibfnamefont {A.}~\bibnamefont
  {Uhlmann}},\ }in\ \href {\doibase 10.1007/978-94-011-2801-8_23} {\emph
  {\bibinfo {booktitle} {Groups Relat. Top.}}},\ \bibinfo {series and number}
  {\bibinfo {number} {1991}}\ (\bibinfo  {publisher} {Springer Netherlands},\
  \bibinfo {address} {Dordrecht},\ \bibinfo {year} {1992})\ pp.\ \bibinfo
  {pages} {267--274}\BibitemShut {NoStop}%
\bibitem [{\citenamefont {Fubini}(1904)}]{Fubini1904}%
  \BibitemOpen
  \bibfield  {author} {\bibinfo {author} {\bibfnamefont {G.}~\bibnamefont
  {Fubini}},\ }\href@noop {} {\bibfield  {journal} {\bibinfo  {journal} {Atti
  Istit. Veneto}\ }\textbf {\bibinfo {volume} {63}},\ \bibinfo {pages} {502}
  (\bibinfo {year} {1904})}\BibitemShut {NoStop}%
\bibitem [{\citenamefont {Study}(1905)}]{Study1905}%
  \BibitemOpen
  \bibfield  {author} {\bibinfo {author} {\bibfnamefont {E.}~\bibnamefont
  {Study}},\ }\href@noop {} {\bibfield  {journal} {\bibinfo  {journal} {Math.
  Ann.}\ }\textbf {\bibinfo {volume} {60}},\ \bibinfo {pages} {321} (\bibinfo
  {year} {1905})}\BibitemShut {NoStop}%
\bibitem [{\citenamefont {Baird}\ and\ \citenamefont
  {Biedenharn}(1963)}]{Baird1963}%
  \BibitemOpen
  \bibfield  {author} {\bibinfo {author} {\bibfnamefont {G.~E.}\ \bibnamefont
  {Baird}}\ and\ \bibinfo {author} {\bibfnamefont {L.~C.}\ \bibnamefont
  {Biedenharn}},\ }\href {\doibase 10.1063/1.1703926} {\bibfield  {journal}
  {\bibinfo  {journal} {J. Math. Phys.}\ }\textbf {\bibinfo {volume} {4}},\
  \bibinfo {pages} {1449} (\bibinfo {year} {1963})}\BibitemShut {NoStop}%
\bibitem [{\citenamefont {Byrd}\ and\ \citenamefont
  {Khaneja}(2003)}]{Byrd2003}%
  \BibitemOpen
  \bibfield  {author} {\bibinfo {author} {\bibfnamefont {M.~S.}\ \bibnamefont
  {Byrd}}\ and\ \bibinfo {author} {\bibfnamefont {N.}~\bibnamefont {Khaneja}},\
  }\href {\doibase 10.1103/PhysRevA.68.062322} {\bibfield  {journal} {\bibinfo
  {journal} {Phys. Rev. A}\ }\textbf {\bibinfo {volume} {68}},\ \bibinfo
  {pages} {062322} (\bibinfo {year} {2003})}\BibitemShut {NoStop}%
\bibitem [{Note1()}]{Note1}%
  \BibitemOpen
  \bibinfo {note} {Some authors have suggested quantifying the driving resource
  independently of the state, for instance in terms of norms of the driving
  Hamiltonian \cite {Uzdin2012,Binder2015, Binder2016, Campaioli2017,
  Deffner2017a, Russell2017}.}\BibitemShut {Stop}%
\bibitem [{\citenamefont {Breuer}\ and\ \citenamefont
  {Petruccione}(2002)}]{Breuer2002}%
  \BibitemOpen
  \bibfield  {author} {\bibinfo {author} {\bibfnamefont {H.-P.}\ \bibnamefont
  {Breuer}}\ and\ \bibinfo {author} {\bibfnamefont {F.~F.}\ \bibnamefont
  {Petruccione}},\ }\href@noop {} {\emph {\bibinfo {title} {{The theory of open
  quantum systems}}}}\ (\bibinfo  {publisher} {Oxford University Press},\
  \bibinfo {year} {2002})\ p.\ \bibinfo {pages} {625}\BibitemShut {NoStop}%
\bibitem [{\citenamefont {Pires}\ \emph {et~al.}(2016)\citenamefont {Pires},
  \citenamefont {Cianciaruso}, \citenamefont {C{\'{e}}leri}, \citenamefont
  {Adesso},\ and\ \citenamefont {Soares-Pinto}}]{Pires2016}%
  \BibitemOpen
  \bibfield  {author} {\bibinfo {author} {\bibfnamefont {D.~P.}\ \bibnamefont
  {Pires}}, \bibinfo {author} {\bibfnamefont {M.}~\bibnamefont {Cianciaruso}},
  \bibinfo {author} {\bibfnamefont {L.~C.}\ \bibnamefont {C{\'{e}}leri}},
  \bibinfo {author} {\bibfnamefont {G.}~\bibnamefont {Adesso}}, \ and\ \bibinfo
  {author} {\bibfnamefont {D.~O.}\ \bibnamefont {Soares-Pinto}},\ }\href
  {\doibase 10.1103/PhysRevX.6.021031} {\bibfield  {journal} {\bibinfo
  {journal} {Phys. Rev. X}\ }\textbf {\bibinfo {volume} {6}},\ \bibinfo {pages}
  {021031} (\bibinfo {year} {2016})}\BibitemShut {NoStop}%
\bibitem [{\citenamefont {Marvian}\ \emph {et~al.}(2016)\citenamefont
  {Marvian}, \citenamefont {Spekkens},\ and\ \citenamefont
  {Zanardi}}]{Marvian2016}%
  \BibitemOpen
  \bibfield  {author} {\bibinfo {author} {\bibfnamefont {I.}~\bibnamefont
  {Marvian}}, \bibinfo {author} {\bibfnamefont {R.~W.}\ \bibnamefont
  {Spekkens}}, \ and\ \bibinfo {author} {\bibfnamefont {P.}~\bibnamefont
  {Zanardi}},\ }\href {\doibase 10.1103/PhysRevA.93.052331} {\bibfield
  {journal} {\bibinfo  {journal} {Phys. Rev. A}\ }\textbf {\bibinfo {volume}
  {93}},\ \bibinfo {pages} {052331} (\bibinfo {year} {2016})}\BibitemShut
  {NoStop}%
\bibitem [{\citenamefont {Mondal}\ and\ \citenamefont
  {Pati}(2016)}]{Mondal2016b}%
  \BibitemOpen
  \bibfield  {author} {\bibinfo {author} {\bibfnamefont {D.}~\bibnamefont
  {Mondal}}\ and\ \bibinfo {author} {\bibfnamefont {A.~K.}\ \bibnamefont
  {Pati}},\ }\href {\doibase 10.1016/J.PHYSLETA.2016.02.018} {\bibfield
  {journal} {\bibinfo  {journal} {Phys. Lett. A}\ }\textbf {\bibinfo {volume}
  {380}},\ \bibinfo {pages} {1395} (\bibinfo {year} {2016})}\BibitemShut
  {NoStop}%
\bibitem [{Note2()}]{Note2}%
  \BibitemOpen
  \bibinfo {note} {In the exceptional case of $N=2$, however, $\Theta $ does
  reduce to the Fubini-Study distance, since the set of all Bloch vectors forms
  a 2-sphere.}\BibitemShut {Stop}%
\bibitem [{Note3()}]{Note3}%
  \BibitemOpen
  \bibinfo {note} {To see this, note that the quantity {$\unhbox \voidb@x \hbox
  {tr}[(\rho _t H_t)^2]=\delimiter "426830A {a}| H_t |{a}\delimiter "526930B
  \delimiter "426830A {a}| H_t|{a}\delimiter "526930B = |\unhbox \voidb@x \hbox
  {tr}(\rho _t H_t)|^2$} for {$\rho =|{a}\delimiter "526930B \delimiter
  "426830A {a}|$}.}\BibitemShut {Stop}%
\bibitem [{\citenamefont {Modi}\ \emph {et~al.}(2016)\citenamefont {Modi},
  \citenamefont {C{\'{e}}leri}, \citenamefont {Thompson},\ and\ \citenamefont
  {Gu}}]{Modi2016}%
  \BibitemOpen
  \bibfield  {author} {\bibinfo {author} {\bibfnamefont {K.}~\bibnamefont
  {Modi}}, \bibinfo {author} {\bibfnamefont {L.~C.}\ \bibnamefont
  {C{\'{e}}leri}}, \bibinfo {author} {\bibfnamefont {J.}~\bibnamefont
  {Thompson}}, \ and\ \bibinfo {author} {\bibfnamefont {M.}~\bibnamefont
  {Gu}},\ }\href {http://arxiv.org/abs/1608.01443} {\  (\bibinfo {year}
  {2016})},\ \Eprint {http://arxiv.org/abs/1608.01443} {arXiv:1608.01443}
  \BibitemShut {NoStop}%
\bibitem [{Note4()}]{Note4}%
  \BibitemOpen
  \bibinfo {note} {The complexity of a matrix multiplication between two $N
  \times N$ matrices is equal to $O(N^{2.373})$, whereas the evaluation of the
  square root of such a matrix has complexity equal to $O(N^3)$ \cite
  {Frommer2010,Davie2013}}\BibitemShut {NoStop}%
\bibitem [{\citenamefont {Ekert}\ \emph {et~al.}(2002)\citenamefont {Ekert},
  \citenamefont {Alves}, \citenamefont {Oi}, \citenamefont {Horodecki},
  \citenamefont {Horodecki},\ and\ \citenamefont {Kwek}}]{Ekert2002}%
  \BibitemOpen
  \bibfield  {author} {\bibinfo {author} {\bibfnamefont {A.~K.}\ \bibnamefont
  {Ekert}}, \bibinfo {author} {\bibfnamefont {C.~M.}\ \bibnamefont {Alves}},
  \bibinfo {author} {\bibfnamefont {D.~K.~L.}\ \bibnamefont {Oi}}, \bibinfo
  {author} {\bibfnamefont {M.}~\bibnamefont {Horodecki}}, \bibinfo {author}
  {\bibfnamefont {P.}~\bibnamefont {Horodecki}}, \ and\ \bibinfo {author}
  {\bibfnamefont {L.~C.}\ \bibnamefont {Kwek}},\ }\href {\doibase
  10.1103/PhysRevLett.88.217901} {\bibfield  {journal} {\bibinfo  {journal}
  {Phys. Rev. Lett.}\ }\textbf {\bibinfo {volume} {88}},\ \bibinfo {pages}
  {217901} (\bibinfo {year} {2002})}\BibitemShut {NoStop}%
\bibitem [{\citenamefont {Keyl}\ and\ \citenamefont {Werner}(2001)}]{Keyl2001}%
  \BibitemOpen
  \bibfield  {author} {\bibinfo {author} {\bibfnamefont {M.}~\bibnamefont
  {Keyl}}\ and\ \bibinfo {author} {\bibfnamefont {R.~F.}\ \bibnamefont
  {Werner}},\ }\href {\doibase 10.1103/PhysRevA.64.052311} {\bibfield
  {journal} {\bibinfo  {journal} {Phys. Rev. A}\ }\textbf {\bibinfo {volume}
  {64}},\ \bibinfo {pages} {052311} (\bibinfo {year} {2001})}\BibitemShut
  {NoStop}%
\bibitem [{Note5()}]{Note5}%
  \BibitemOpen
  \bibinfo {note} {A similar experimental set up could be used to evaluate the
  bound derived by authors in Ref.~\cite {Mondal2016b}.}\BibitemShut {Stop}%
\bibitem [{\citenamefont {Uzdin}\ \emph {et~al.}(2012)\citenamefont {Uzdin},
  \citenamefont {G{\"{u}}nther}, \citenamefont {Rahav},\ and\ \citenamefont
  {Moiseyev}}]{Uzdin2012}%
  \BibitemOpen
  \bibfield  {author} {\bibinfo {author} {\bibfnamefont {R.}~\bibnamefont
  {Uzdin}}, \bibinfo {author} {\bibfnamefont {U.}~\bibnamefont
  {G{\"{u}}nther}}, \bibinfo {author} {\bibfnamefont {S.}~\bibnamefont
  {Rahav}}, \ and\ \bibinfo {author} {\bibfnamefont {N.}~\bibnamefont
  {Moiseyev}},\ }\href {\doibase 10.1088/1751-8113/45/41/415304} {\bibfield
  {journal} {\bibinfo  {journal} {Journal of Physics A: Mathematical and
  Theoretical}\ }\textbf {\bibinfo {volume} {45}},\ \bibinfo {pages} {415304}
  (\bibinfo {year} {2012})}\BibitemShut {NoStop}%
\bibitem [{\citenamefont {Deffner}(2017)}]{Deffner2017a}%
  \BibitemOpen
  \bibfield  {author} {\bibinfo {author} {\bibfnamefont {S.}~\bibnamefont
  {Deffner}},\ }\href {https://arxiv.org/abs/1704.03357} {\  (\bibinfo {year}
  {2017})},\ \Eprint {http://arxiv.org/abs/1704.03357} {arXiv:1704.03357}
  \BibitemShut {NoStop}%
\bibitem [{\citenamefont {Russell}\ and\ \citenamefont
  {Stepney}(2017)}]{Russell2017}%
  \BibitemOpen
  \bibfield  {author} {\bibinfo {author} {\bibfnamefont {B.}~\bibnamefont
  {Russell}}\ and\ \bibinfo {author} {\bibfnamefont {S.}~\bibnamefont
  {Stepney}},\ }\href {\doibase 10.1142/S0129054117500204} {\bibfield
  {journal} {\bibinfo  {journal} {Int. J. Found. Comput. Sci.}\ }\textbf
  {\bibinfo {volume} {28}},\ \bibinfo {pages} {321} (\bibinfo {year}
  {2017})}\BibitemShut {NoStop}%
\bibitem [{\citenamefont {Frommer}\ and\ \citenamefont
  {Hashemi}(2010)}]{Frommer2010}%
  \BibitemOpen
  \bibfield  {author} {\bibinfo {author} {\bibfnamefont {A.}~\bibnamefont
  {Frommer}}\ and\ \bibinfo {author} {\bibfnamefont {B.}~\bibnamefont
  {Hashemi}},\ }\href {\doibase 10.1137/090757058} {\bibfield  {journal}
  {\bibinfo  {journal} {SIAM J. Matrix Anal. Appl.}\ }\textbf {\bibinfo
  {volume} {31}},\ \bibinfo {pages} {1279} (\bibinfo {year}
  {2010})}\BibitemShut {NoStop}%
\bibitem [{\citenamefont {Davie}\ and\ \citenamefont
  {Stothers}(2013)}]{Davie2013}%
  \BibitemOpen
  \bibfield  {author} {\bibinfo {author} {\bibfnamefont {A.~M.}\ \bibnamefont
  {Davie}}\ and\ \bibinfo {author} {\bibfnamefont {A.~J.}\ \bibnamefont
  {Stothers}},\ }\href {\doibase 10.1017/S0308210511001648} {\bibfield
  {journal} {\bibinfo  {journal} {Proc. R. Soc. Edinburgh Sect. A Math.}\
  }\textbf {\bibinfo {volume} {143}},\ \bibinfo {pages} {351} (\bibinfo {year}
  {2013})}\BibitemShut {NoStop}%
\end{thebibliography}%

\pagebreak
\clearpage
\pagebreak
\begin{center}
\textbf{\large Supplemental Material}
\end{center}
\section{Proof of Remark~\ref{r:pure_limit}}
\label{s:pure_limit}
\noindent
Since bound $T_\Theta$ expressed in Eq.~\eqref{eq:qsl_gba} is clearly different from the QSL for pure states, we show that they coincide when $N=2$. 
First, we show that $\Theta$ concides with the Fubini-Study distance for the case of 2-dimensional systems.
Let $\rho = \ket{a}\bra{a}$ and $\sigma=\ket{b}\bra{b}$ for two pure qubit states $\ket{a},\ket{b}$, with associated Bloch vectors $\bm{a}, \bm{b}$. Let us fix the basis such that $\ket{a}=\ket{(\varphi,\theta)}$ where $\ket{(\varphi,\theta)}$ relative to Bloch vector $\bm{a} = (\cos\varphi\sin\theta,\sin\varphi\sin\theta,\cos\theta)$ with $\varphi\in[0,2\pi], \; \theta\in[0,\pi]$, and where $\ket{b}=\ket{(0,0)}$ is aligned with the $\hat{z}$-axis. State $\ket{a} = \cos (\theta/2) \ket{b} + e^{i\varphi} \sin(\theta/2)\ket{\bar{b}}$, where $\ket{\bar{b}} = \ket{0,\pi}$ is orthogonal to $\ket{b}$. The Fubini-Study distance $d(\ket{a},\ket{b}) = \arccos{|\braket{a|b}|}=\theta/2$, while $\Theta(\rho,\sigma) = \theta$, thus the two distances are identical up to a factor of $1/2$. For pure qubit ($N=2$) states $Q_\Theta = 2 \Delta E$, since $\tr[\rho^2]=1$, therefore $T_\Theta =  \theta/\Delta E$. Hence, $T_\Theta$ and the Mandelstam-Tamm bound coincide for qubits. For dimension $N>2$, $Q_\Theta$ reduces neither to the standard deviation, nor to the average energy, while $\Theta$ does not become the Fubini-Study distance.

\section{Proof of Theorem~\ref{th:qsl_tilde}}
\label{s:proof_phi}
\noindent
First, we prove that $\Phi$ is a distance that reduces to the Fubini-Study distance for the case of pure states. $\Phi(\rho,\sigma)\geq 0$ since $\textrm{tr}[\rho\sigma]=\sum_{a,b}\lambda_a\lambda_b|\braket{r_a|s_b}|^2$ is positive and always smaller or equal to $\mathrm{tr}[\rho^2] = \mathrm{tr}[\sigma^2] $. This can be proved using the Hilbert-Schmitdt distance $D_2^2(A,B):=\mathrm{tr}[(A-B)(A^\dagger-B^\dagger)]/2\geq0$ for $\rho$ and $\sigma$, obtaining $\mathrm{tr}[\rho^2]-\mathrm{tr}[\rho\sigma]\geq 0$. $\Phi(\rho,\rho) =0 $, and if $\Phi$ is zero $\mathrm{tr}[\rho\sigma] = \mathrm{tr}[\rho^2] = \mathrm{tr}[\sigma^2]$ therefore $\sigma = \rho$. Symmetry holds due to cyclicity of the trace and because $\rho$ and $\sigma$ have the same purity $\mathrm{tr}[\rho^2]$. $\Phi(\rho,\sigma)=\arccos(\sqrt{(1-(N-1)\lVert \bm{r}\rVert_2^2 \hat{\bm{r}}\cdot\hat{\bm{s}})/(1-(N-1)\lVert \bm{r}\rVert_2^2 )})$ is monotonic in $\hat{\bm{r}}\cdot\hat{\bm{s}}$, thus respects the triangle inequality.  Ergo, $\Phi$ is a distance on the space of those states that can be unitarily connected. For the case of pure states, $\Phi$ reduces to the Fubini-Study distance, since, for $\rho =\ket{\psi}\bra{\psi}$ and $\sigma =\ket{\phi}\bra{\phi}$, $\sqrt{\mathrm{tr}[\rho\sigma]} = |\braket{\psi|\phi}|$ and $\mathrm{tr}[\rho^2]=1$.

The proof of Eq.~\eqref{eq:qsl_tilde} is identical to that for Eq.~\eqref{eq:qsl_gba}, expect for the fact that $\Phi(\rho_t,\rho_{t+dt}) = \arccos (\sqrt{1-dt^2(\tr[\rho_t^2 H_t^2]-\tr[(\rho_t H_t)^2])/\tr[\rho_t^2]}$). We expand $\arccos(\sqrt{1-c}) \sim \arccos(1-c/2) \sqrt{c}+\mathcal{O}(c)$ for small $c>0$, obtaining
$\int_0^T\Phi(\rho_t,\rho_{t+dt}) =  T Q_\Phi $ which leads to $\Phi(\rho,\sigma) \leq Q_\Phi \ T$. Thus, Eq.~\eqref{eq:qsl_tilde} holds.

\section{Quantum Speed Limits for Qutrits}
\label{s:qutrits}
\begin{figure}[h]
    \centering
    \includegraphics[width=0.5\textwidth]{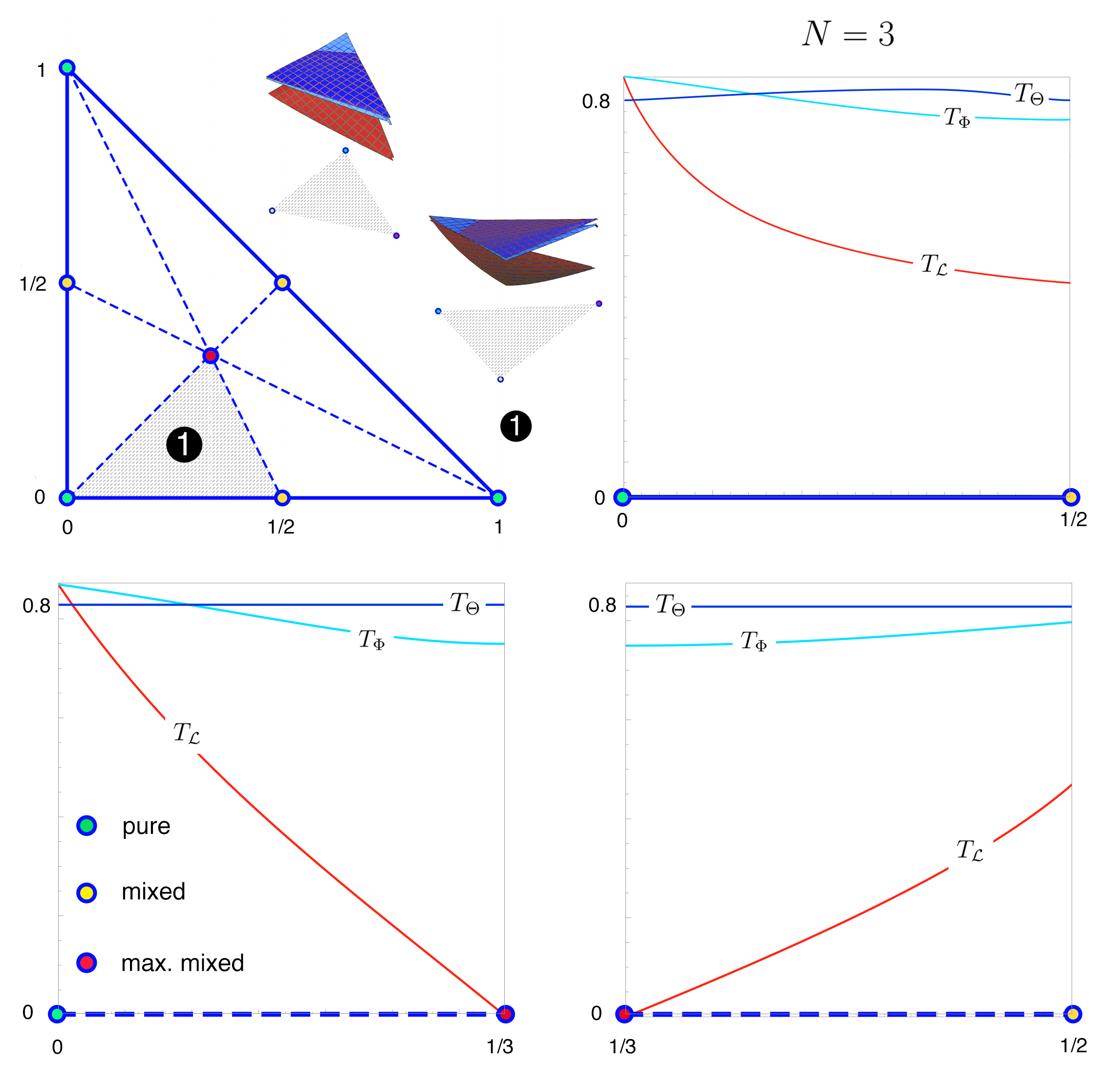}
    \caption{\emph{Example:} bounds $T_{\mathcal{L}}$ (Eq.~\eqref{eq:QSL_mixed}, red line), $T_\Phi$ (Eq.~\eqref{eq:qsl_tilde}, light blue line),
    and $T_\Theta$ (Eq.~\eqref{eq:qsl_gba}, blue line), as a function of the eigenvalues $\lambda_1,\lambda_2,\lambda_3$, for a specific choice of mixed
    qutrit state $\rho$ and Hamiltonian $H$, for driving $\rho\to\sigma=O\rho O^\dagger$, with $O=\exp[-i H]$. Green vertices correspond to pure states, yellow and red vertices to maximally mixed states of rank 2 and 1, respectively. 
    The insets in the first graph show a 3D-rendering of all three speed limits in the whole of region \circled{1}; the other three graphs each show the the speed limits along one of the edges of that region.}
    \label{fig:qutrits}
\end{figure}

\noindent
In this section we study the new bounds in Eqs.~\eqref{eq:qsl_gba} and~\eqref{eq:qsl_tilde} for the case of qutrits, and we compare them to the one in Eq.~\eqref{eq:QSL_mixed}. The spectrum $\Lambda=\{\lambda_1,\lambda_2,\lambda_3\}$ can be represented with the standard 2-simplex $\Delta_2$ (equilateral triangle), or by its projection onto the plane defined by $\lambda_1$ and $\lambda_2$, since the third component $\lambda_3$ has to be equal to $1-\lambda_1-\lambda_2$, and $\lambda_1+\lambda_2\leq1$. Within this space, we only need to consider the region \circled{1} given by 
$0\leq \lambda_1\leq 1/2$, $\lambda_2\leq\lambda_1$ and $\lambda_2\leq 1-2\lambda_2$ (see Fig.~\ref{fig:qutrits})~\cite{Bengtsson2008}. This region is determined by the three vertices $(0,0,1)$ (green dot), $(1/2,0,1/2)$ (yellow dot), and $(1/3,1/3,1/3)$ (red dot), that correspond to a pure state, a mixed state with two identical eigenvalues equal to $1/2$, and the maximally mixed state, respectively, and delimited by the segments that connect these vertices. These segments are characterized by two kinds of degeneracy: the one that connects the pure state to the mixed state associated with $(1/2,0,1/2)$ (given by $(\lambda_1,0,1-\lambda_1)$, solid blue line), is composed of fully non-degenerate mixed states, while the other two segments (dashed blue lines) contain mixed states with two identical eigenvalues (the same degeneracy structure as for pure states).

We generated Haar random states and Hamiltonians, and then studied the three different bounds in the region \circled{1} (see Fig.~\ref{fig:qutrits}). As described in the main text, $T_\Phi = T_{\mathcal{L}} > T_\Theta$ at the pure vertex. The bound $T_\Theta$ is constant along the dashed lines, while it can vary continuously along the solid solid blue line. For all of the generated Hamiltonians, $\max[T_\Theta,T_\Phi]-T_{\mathcal{L}}>0$. 

\end{document}